# Point Defects in Double Helix Induced by Interaction of Silver Nanoparticles with DNA

Vasil G. Bregadze[*], Zaza G. Melikishvili[**], Tamar G. Giorgadze[*], Jamlet R. Monaselidze[*],

Zaza V. Jaliashvili[**], Temur B. Khuskivadze[*]

* Ivane Javakhishvili Tbilisi State University, Andronikashvili Institute of Physics, 6 Tamarashvili Str. 0177 Tbilisi, Georgia;
**Georgian Technical University, Institute of Cybernetics. 5, Euli Str., 0186, Tbilisi, Georgia.

**Abstract**. Interaction of DNA-silver nanoparticles (AgNPs) complexes with $H_3O^+$, $Cu^{2+}$ and $Cl^-$ has been studied by spectro-photometric, spectro-fluorimetric and differential scanning micro calorimetric methods. It is shown that DNA is a catalyst in redox reactions taking place in AgNPs adsorbed on its surface. We also demonstrate that $Ag^+$ ions that are freed after corrosion of nanoparticles show absorption into the inner part of DNA double helix, i.e. they make the so-called cross-links between complementary base pairs of DNA. The cross-links present point defects of DNA which leads *in-vivo* to cell death.

**Keywords:** Silver nanoparticles, silver ions, DNA-intercalator complex, optical spectroscopy of intramolecular interactions.

**Corresponding author**: Vasil G. Bregadze vbregadze@gmail.com, v.bregadze@aiphysics.ge

## Introduction

It has become more evident lately that in a lot of processes where DNA is an adsorbent of small ligands it is the catalyst of the processes. For instance, metal induced double proton transfer in GC pairs [1]; corrosion of silver nanoparticles (AgNPs) [2]; diffusion of AgNPs on DNA [3]; energy transfer from acridine orange (AO) to ethidium bromide (EB) [4]; reduction of metal ions [5]. Besides, application of metal nanoparticles, particularly gold ones, in cancer photo-chemo and photo-thermo therapy is also very promising [6-10]. Silver nanoparticles are considerably rarely used for the purpose because one of their substantial properties is their stability in solutions [11].

The goal of the present contribution is the study of AgNPs interactions with DNA in solutions by spectroscopic and thermodynamic methods and the effect of $H_3O^+$ and $Cu^{2+}$ ions on the process.

## Materials and Methods

- Colloidal silver suspension in distilled water was made of silver nanoparticles (AgNPs) of 1-2 nm size (DDS Inc., D/B/A/, Amino Acid & Botanical Supply).
- Calf thymus DNA produced by Sigma was dissolved in 0.01M $NaNO_3$ solution – background electrolyte, pH~6.0. Nucleotide concentration was evaluated by UV absorption spectrophotometer. Molecular extinction factor $ɛ = 6600$ $cm^{-1}$ $M^{-1}$ (P), $\lambda = 260$ nm.
- **Intercalators.** Acridine orange (AO) was purchased from 'Sigma'. The concentration of the dye was determined colorimetrically at the isobestic point of the monomer-dimer system ($\lambda = 470$ nm) using the molar extinction coefficients ($ɛ = 43\ 300$ $cm^{-1}$ $M^{-1}$). Ethidium bromide (EB) was also purchased from 'Sigma'. The concentration of the dye was determined colorimetrically ($ɛ = 5600$ $cm^{-1}$ $M^{-1}$ at $\lambda = 480$ nm).

- **Ions.** We used chemically pure chlorid of Cu and and also especially pure NaCl. Bidistillate served as a solvent. In tests with $Ag^+$ ions chemically pure salts $AgNO_3$ and $NaNO_3$ were used as background electrolytes.
  - Registration of absorption spectra was carried out by optical fiber spectrometer **AvaSpec ULS 2048 – USB 2.**

**Results and Discussion**

Fig.1 shows absorption spectra of: a) AgNPs suspended in the solution of $NaNO_3$ ($10^{-2}$ M); b) AgNPs in the complex with DNA and c) AgNPs in the complex with denaturized DNA (DNA solution had been incubated at $100^0C$ for 15 min and then it was suddenly cooled in ice bath). Fig. 2 presents first derivatives of absorption spectra of AgNPs and AgNPs in the complex with DNA.

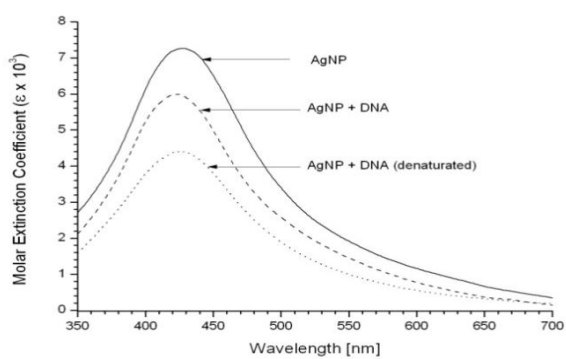

Fig.1 Absorption spectra of: AgNPs suspended in the solution of $NaNO_3$ ($10^{-2}$M); AgNPs in the complex with DNA and AgNPs in the complex with denaturized DNA (DNA solution had been incubated at $100^0C$ for 15 min and then it was suddenly cooled in ice bath). [AgNPs]-$0.72 \cdot 10^{-4}$M ($Ag^0$), [DNA]-$1.6 \cdot 10^{-4}$ M(P)

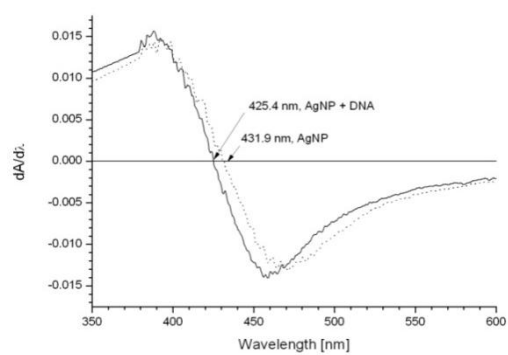

Fig.2 First derivatives of absorption spectra of AgNPs and AgNPs in the complex with DNA

The analysis of Figs. 1 and 2 shows that at DNA interaction with AgNPs a short-wave shift of AgNPs absorption band (6.5nm) takes place. Besides, there can be observed a 20% hypochromic effect. The short-wave shift points out that at interaction with DNA there is a kind of loosening of interaction between silver atoms inside the AgNPs. Decrease of the intensity of the absorption band is due to partial corrosion of AgNPs in the presence of DNA. Strong effect on AgNPs is rendered when they are in the complex with denaturized DNA, i.e. DNA with partially unfolded regions of double helix. It can be explained by the ability of DNA to adsorb $H_3O^+$ (pK~4) and metal ions of the first transition row $M^{2+}$ (pK=4–6) [12]. It should be noted that at study of the interaction of transition metal ions with DNA non-buffer solutions are usually used. As a rule pH of the solutions is 5.5–6. At these pH values the concentration of $H_3O^+$ in the solution is satisfactory for the attacks at DNA because $H_3O^+$ adsorption on DNA is of mobile character [5]. Thus, the surface of the DNA can adsorb both AgNPs and $H_3O^+$ and serve as a catalyst in AgNPs oxidation reaction. It is necessary to underline that AgNPs used in the experiment have the size of 1 -2 nm and, consequently, their concentration is at least two orders less than the one for atoms which constitute the particles. Assuming that all nanoparticles of the solution are adsorbed on DNA and they interact from the side of double helix major groove and their average size is 1.5nm, we can conclude that the average distance between AgNPs adsorbed on DNA is nearly 60 nm (1 particle for about 170 base pairs). As soon as $H_3O^+$ reduction starts (as a result of AgNPs oxidation) the water in the solution will dissociate to $OH^-$ and $H^+$ supplying new $H_3O^+$ until new dynamic balance of the system is achieved. $Ag^+$ ions which are created during the reaction have high stability constant with DNA (pK≥10.8 [5]). As far as in 1969 Wilhelm and Daune [13] showed that $Ag^+$ ions make two kinds of intra-spherical complexes with G-C DNA pairs: chelate $N_7G - O_6G$ and intra-strand linear complex between $N_1G$ and $N_3C$, so-called cross-link. The authors [13] believe that at making the complex of the second type $H_3O^+$ is released from DNA guanine into the solution. It is an additional mechanism of $H_3O^+$ formation in solution. As formation of the second type complex depends on DNA dynamic characteristics, i.e. on frequency of unfolding the base pairs, the possibility of the process will be much higher in case when DNA is denatured. We could observe it in our experiment (see Fig.1).

$Cu^{2+}$ ions ($CuCl_2$) should have higher oxidizing ability for AgNPs adsorbed on DNA than $H_3O^+$. First of all they adsorb on DNA with the stability constant pK~5–6 [12,1] and possess higher than $H_3O^+$ affinity to electron [14-15], though they do not possess mobile adsorption and besides, they are added to the solution having concentration of $10^{-4}$M. Fig. 3 presents $CuCl_2$ effect on absorption spectra of AgNPs-DNA complex against the sequence of adding $CuCl_2$. Fig. 4 shows the effect of various $CuCl_2$ concentrations.

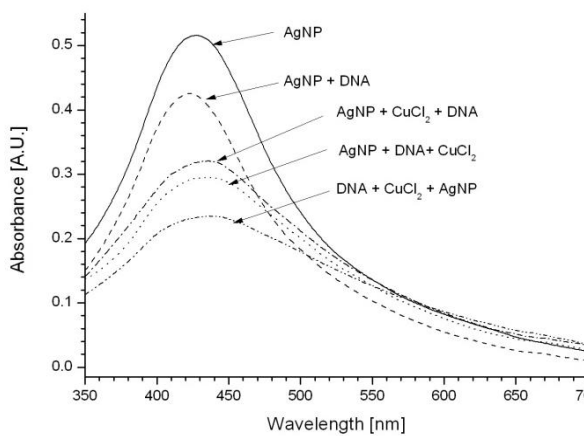
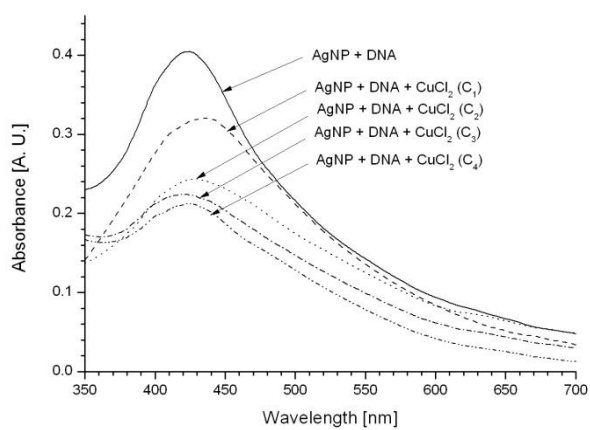

Fig.3 $CuCl_2$ effect on absorbtion spectra of AgNPs-DNA complex against the sequence of adding $CuCl_2$. [AgNPs]-0.72·$10^{-4}$ M($Ag^0$), [DNA]-1.6·$10^{-4}$ M(P), [$CuCl_2$]- 0.75·$10^{-4}$ M, [$NaNO_3$]-$10^{-2}$M

Fig.4 Effect of various $CuCl_2$ concentrations on absorbtion spectra of AgNPs-DNA complexes. [AgNPs]-0.72·$10^{-4}$ M ($Ag^0$), [DNA]-1.6·$10^{-4}$ M(P), [$CuCl_2$]- 0.75·$10^{-4}$ M($C_1$), [$CuCl_2$]- 1.52·$10^{-4}$ M($C_2$), [$CuCl_2$]- 2.3·$10^{-4}$ M($C_3$), [$CuCl_2$]- 3.0·$10^{-4}$ M($C_4$), [$NaNO_3$]-$10^{-2}$M

To exclude the effect of Cl⁻ ions a special experiment with NaCl has been carried out (see Fig.5). It is known that Cl⁻ ions can activate precipitation of $Ag^+$ but as we see DNA blocks it, moreover Cl⁻ ions reduce corrosion of AgNPs caused by its interaction with DNA.

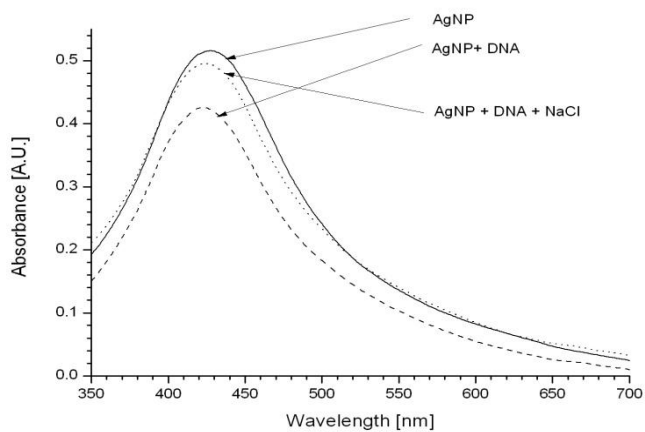

Fig.5 Effect of Cl⁻ ions on absorbtion spectra of AgNPs-DNA complexes.[AgNPs]-$0.72 \cdot 10^{-4}$ M ($Ag^0$), [DNA]-$1.6 \cdot 10^{-4}$ M(P), [NaCl]-$0.75 \cdot 10^{-4}$ M, [NaNO$_3$]-$10^{-2}$M

For better justification of our approach to the corrosion mechanism of AgNPs adsorbed on DNA we have conducted a comparative fluorimetric analysis of AgNPs and $Ag^+$ (AgNO$_3$) impact on efficiency of nonradiative energy transfer of electron excitation in the pair of dyes which are intercalated in DNA – ethidium bromide (acceptor) and acridine orange (donor). The effectiveness of the transfer for intercalaters much depends on the distance between them ($1/R^6$). Stability constant of $Ag^+$ ions complex with DNA is at least 5 orders higher than stability constant of AO and EB complexes with DNA. At $Ag^+$ interaction with DNA the intercalated dyes are due to subject their binding places to $Ag^+$ ions and tighten in DNA up to their ejection into the solution [4,5]. The results of the above fluorimetric analysis are presented on Fig.6 and 7. Concentrations of DNA ($2.8 \times 10^{-4}$M (P) or $1.4 \times 10^{-4}$M base pairs) and intercalaters ($0.14 \times 10^{-4}$ M) have been chosen in such a way that on average intercalaters were located after each 5 DNA base pairs or the distance between them was 2 nm (0.34 nm x 6 for DNA B form). Fluorescence spectra presented on the both pictures demonstrate that before $Ag^+$ ions and AgNPs were added into the solution the efficiency of energy transfer from AO to EB was low. After addition of small amount of $Ag^+$ and AgNPs energy transfer efficiency notably increased and AO fluorescence was quenched at the expense of EB inflammation.

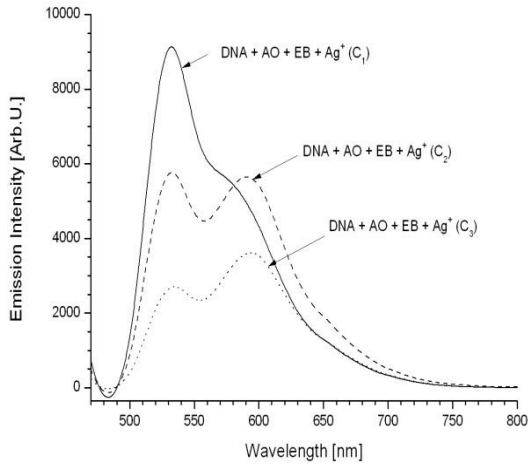 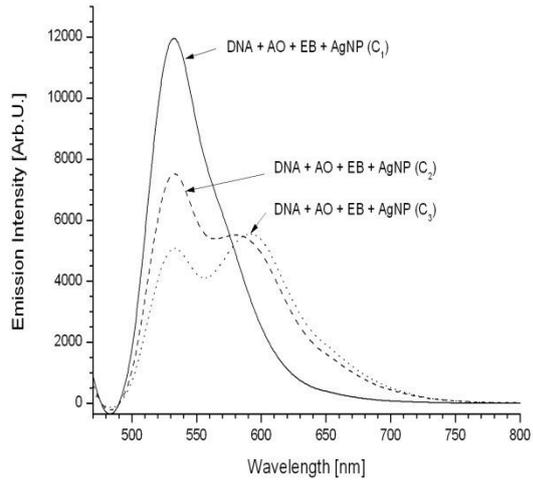

Fig.6 Quenching of fluorescence by $Ag^+$ ions in DNA-AO-EB complex. [DNA]-$2.8 \cdot 10^{-4}$ M (P), [AO]-$0.14 \cdot 10^{-4}$ M, [EB]- $0.14 \cdot 10^{-4}$ M, ), [$Ag^+$]-0 ($C_1$), [$Ag^+$]-$6.0 \cdot 10^{-6}$ M ($C_2$), [$Ag^+$]-$30.0 \cdot 10^{-6}$ M ($C_3$), [$NaNO_3$]-$10^{-2}$ M. $\lambda_{ex}$ = 460 nm.

Fig.7 Quenching of fluorescence by AgNPs in DNA-AO-EB complex. [DNA]-$2.8 \cdot 10^{-4}$ M (P), [AO]-$0.14 \cdot 10^{-4}$ M, [EB]- $0.14 \cdot 10^{-4}$ M, ), [AgNPs]-0 ($C_1$), [AgNPs]-$6.0 \cdot 10^{-6}$ M ($C_2$), [AgNPs]-$18.0 \cdot 10^{-6}$ M ($C_3$), [$NaNO_3$]-$10^{-2}$M. $\lambda_{ex}$ = 460 nm.

Besides, it is worth mentioning that the experiments where original differential scanning micro-calorimeter was applied [16] prove our conclusion concerning AgNPs corrosion in the complex with DNA. $Ag^+$ ions as well as AgNPs adsorbed on DNA increase melting temperature of calf thymus DNA and moreover its cooling to $20^0$C reduces significant part of DNA double helix. It has been proved by repetitive heating of the complexes.

**Acknowledgments:**

The work was partly supported by the Grant No GNSF/ST09_508_2-230.

## References

1. Bregadze V., Gelagutashvili E., and Tsakadze K. Thermodynamic Models Of Metal Ion-DNA Interactions. In: Metal Complexes-DNA interactions. N. Hadjiliadis and E. Sletten, eds. Chichester: Blackwell Publishing Ltd, 2009, 31-53.
2. V.Bregadze, S. Melikishvili, Z. Melikishvili, G. Petriashvili, Nanochemistry and Nanotechnologies Proceedings of Papers of the First International Conference March 23-24, 2010, Tbilisi Georgia, Publishing House "UNIVERSAL" Tbilisi 2011, 136-140.
3. See contribution "Photodiffusion of Silver Nanoparticles on DNA and Medicine" in the present voliume.
4. Bregadze, V.G.; Chkhaberidze, J.G. and Khutsishvili, I.G., Effect of metal ions on the fluorescence of dyes bound to DNA. In: Metal Ions in Biological Systems. A. Sigel and H.Sigel, eds. New York, Basel: Marcel Dekker Inc., 1996, Vol.33, 253-266.
5. Bregadze V.G., Khutsishvili I.G., Chkhaberidze J.G., Sologashvili K. Inorganica Chimica Acta, 2002, V. 339, 145-159.
6. Niidome T, Shiotani A, Akiyame Y, Ohga A, Nose K, Pissuwan D, Niidome Y. Yakugaku Zasshi. 2010 Dec; 130(12):1671-7. Review. Japanese.
7. Khlebtson N, Dykman L. Chem Soc Rev. 2011 Mar; 40(3):1647-71. Epub 2010 Nov 16. Review.
8. Sekhon BS, Kamboj SR. Nanomedicine. 2010 Oct; 6(5):612-8. Epub 2010 Apr 22. Review.
9. Ratto F, Matteini P, Centi S, Rossi F, Pini R. J Biophotonics. 2011 Jan;4(1-2):64-73. Doi: 10.1002/jbio.201000002.Epub 2010 Mar 1. Review.
10. Pissuwan D, Valenzuela S, Cortie MB. Biotechnol Genet Eng Rev. 2008;25:93-112. Review.
11. The effect of aggregation on optical properties: Data of firm nanoComposix, Silver Nanoparticles: Optical Properties. www.nanocomposix.com.
12. Bregadze, V.G. Metal Ion Interactions with DNA: Consideration on Structure, Stability, and Effects from Metal Ion Binding. In: Metal Ions in Biological Systems. A. Sigel and H. Sigel, eds. New York: Marcel Dekker, 1996, Vol.32, 453-474.
13. Wilhelm F. X, et Daune M.,Biopolymers, Vol. 8, 1 pp. 121-137 (1969).
14. Hart, E.J.; Anbar, M.; The Hydrated Electron, Atomizdat, Moscow, 1973, 280.
15. Bregadze, V.G.; Int. J. Quant. Chem. 1980, 17, 1213-1219.
16. J. Monaselidze, M. Abuladze, N. Asatiani, E. Kiziria, Sh.Barbakadze, G. Majagaladze, M. Iobadze, L. Tabatadze, Hoi-Ying Holman, N. Sapojnikova. Thermochemia Acta, 2006, 441, pp. 8-15

**Resume**

1. Interaction of AgNPs with DNA and the effect of $H_3O^+$ and $Cu^{2+}$ ions on the process have been studied by spectro-photometry in visible spectra region. It is shown that AgNPs being adsorbed on the surface of DNA double helix from the side of the big groove activate hypsochromic shift by 6.5 nm and hypochromic effect of about 20%. Hypsochromic shift points out to some loosening of interaction between silver atoms inside AgNPs as a result of attraction of surface atoms from silver nanoparticles by chelate groups $N_7 - O_6$ of DNA guanine. Decrease of absorption in its turn demonstrates AgNPs partial corrosion, i.e. $Ag^0$ oxidation to $Ag^+$ by means of $H_3O^+$ and $Cu^{2+}$ ions which are also adsorbed on DNA. So, DNA can be considered a catalyst in redox reactions in ternary complexes DNA + AgNPs + $H_3O^+$ or $Cu^{2+}$.
2. AgNPs corrosion on DNA is proved by spectro-fluorimetric and micro-calorimetric comparative analysis of AgNPs and $Ag^+$ ($AgNO_3$) effect on the increase of:
a) Efficiency of nonradiative energy transfer in intercalating dye pair AO –EB in ternary complexes DNA –AO –EB;
b) DNA melting temperature and, what is essential, reduction of significant part of DNA helix after cooling to $20^0C$. It has been proved by repeated heating of the complexes.